\documentclass[aps,prd,twocolumn,tightenlines]{revtex4-1}
\usepackage{graphicx}% Include figure files
\usepackage{dcolumn}% Align table columns on decimal point
\usepackage{bm}% bold math
\usepackage{amsmath}
\usepackage{multirow}

\begin{document}

\newcommand{\mevcc}{\!\mathrm{MeV}\!/c^2}
\newcommand{\mevc}{\!\mathrm{MeV}/\!c}
\newcommand{\mev}{\!\mathrm{MeV}}
\newcommand{\gevcc}{\!\mathrm{GeV}/\!c^2}
\newcommand{\gevc}{\!\mathrm{GeV}/\!c}
\newcommand{\gev}{\!\mathrm{GeV}}
\title{High precision measurement of the masses of the $D^0$ and $K_S$ mesons}

\author{A.~Tomaradze}
\author{S.~Dobbs}
\author{T.~Xiao}
\author{Kamal~K.~Seth}
\affiliation{Northwestern University, Evanston, Illinois 60208, USA}

\author{G. Bonvicini} 
\affiliation{Wayne State University, Detroit, Michigan 48202, USA} 

\date{\today}

%\linespread{1.5}

\begin{abstract} 
Using 580~pb$^{-1}$ of $e^+e^-$ annihilation data taken with the CLEO--c detector at $\psi(3770)$,
 the decay $D^0(\overline{D}^0)\to K^\pm\pi^\mp \pi^+\pi^-$ has been studied to make the highest precision measurement of $D^0$ mass, $M(D^0)=1864.845\pm0.025\pm0.022\pm0.053$~MeV, where the first error is statistical, the second error is systematic, and the third error is due to uncertainty in kaon masses.
As an intermediate step of the present investigation the mass of the $K_S$ meson has been measured to be 
 $M(K_S)=497.607\pm0.007\pm0.015$~MeV. Both $M(D^0)$ and $M(K_S)$ are the most precise single measurements of the masses of these mesons.

\end{abstract}

\maketitle

%\linespread{1.5}

The $D^0$ meson, the ground state of the charm meson family, 
and the $K_S$  meson, the ground store of 
the strange meson family, occupy an important place in 
hadron spectroscopy, and precision determination 
of their masses is of particular importance. Not only 
do the masses of $K_S$ and $D^0$ mesons provide 
precision calibration standards for masses and mass differences
below 2.5 GeV as $M(J/\psi)$ and $M(\psi(2S))$
do in the 3--4 GeV mass region \cite{kedr1,kedr}, but precision 
determination of  $M(D^0)$  is of 
crucial importance in time--dependent analyses of    
$D^0$--$\bar{D^0}$  mixing and  CP violation 
\cite{asner,babarcp,lhcbcp}. Recently, many observations of mesons that 
do not conveniently fit in the conventional $|q\bar{q}>$  meson families 
have been reported, and several of these are conjectured to be weakly 
bound hadronic molecules of $D$ and  $D_S$  mesons \cite{brambilla}. 
The most famous of these 'exotics' is the X(3872) meson which can be   
modeled as a $\bar{D^{0}}D^{*0}$  molecule. The small binding 
energy of X(3872)  requires a precision determination of the masses 
of $D^{0}$ and $D^{*0}$ mesons \cite{kam,d0pub}. In this paper we present 
results for the highest precision measurement of $M(D^{0})$. As an 
intermediate step in our analysis, we have also made a 
precision measurement of $M(K_S)$.

We had earlier \cite{d0pub} reported the measurement of $M(D^0)$  
 using  280 pb$^{-1}$ of CLEO--c data 
taken at the $\psi(3770)$. 
We reported $M(D^0)=1864.847\pm0.150\pm0.095$~MeV (throughout this paper the first error is statistical, and the second error is systematic), using the decay $D^0\to K_S\phi$, $\phi\to K^+K^-$, $K_S\to {\pi}^+ {\pi}^-$, which has the overall branching fraction $\mathcal{B}=1.4\times10^{-3}$. The measurement was based on $319\pm18$ events. Recently, the LHCb Collaboration has reported $M(D^0)=1864.75\pm0.15\pm0.11$~MeV~\cite{lhcb} based on $4608\pm89$ events in the decay $D^0\to K^+K^-\pi^+\pi^-$, which has $\mathcal{B}=2.4\times10^{-3}$~\cite{pdg}, and $849\pm36$ events in the decay $D^0\to K^+K^-K^-\pi^+$, which has $\mathcal{B}=2.2\times10^{-4}$~\cite{pdg}. Also, BaBar has reported 
$M(D^0)=1864.841\pm0.048\pm0.063$~MeV~\cite{babar} based on $4345\pm70$ events observed in the $D^0\to K^+K^-K^-\pi^+$ decay. The goal of our present measurement is to determine the mass of $D^0$ with an overall precision three times better than our previous measurement, i.e., $\sim60$~keV. To minimize statistical errors we choose to study the most prolific charged particle decay, $D^0\to K^-\pi^+\pi^-\pi^+$ ($K3\pi$) (Throughout this paper inclusion of charge conjugate decays is implied), which has a branching fraction $\mathcal{B}=8.1\times10^{-2}$~\cite{pdg}, sixty times that in our previous measurement, and $\sim370$ times larger than that for the $D^0\to3K\pi$ decay used by BaBar and LHCb. To obtain the best energy calibration for charged hadrons, we analyze the decay $\psi(2S)\to J/\psi\pi^+\pi^-$, and anchor our energy calibration to the high precision measurement of the mass of $J/\psi$,
$M(J/\psi)=3096.917\pm0.010\pm0.007$~MeV~\cite{kedr1}, and mass 
of $\psi(2S)$, 
$M(\psi(2S))=3686.114\pm0.007\pm0.011_{-0.012}^{+0.002}$~MeV~\cite{kedr},
made by the KEDR Collaboration at Novosibirsk using the resonance depolarization technique.

We use data taken with the CLEO--c detector, 580 $\mathrm{pb^{-1}}$ of $e^+e^-$ annihilation at $\psi(3770)$, $\sqrt{s}=3770$ MeV, twice as much as in our previous measurements to determine $D^0$ mass, and 49 $\mathrm{pb^{-1}}$ of data taken at $\psi(2S)$, $\sqrt{s}=3686$ MeV to fine tune the CLEO--c solenoid magnetic field. The CLEO--c detector has been described in detail elsewhere~\cite{cleoc}. Briefly, it consists of a CsI(Tl) electromagnetic calorimeter, an inner vertex drift chamber, a central drift chamber, and a ring imaging Cherenkov (RICH) detector, all inside a superconducting solenoid magnet providing a nominal 1.0 Tesla magnetic field.  For the present measurements, the important components are the drift chambers, which provide a coverage of 93\% of $4\pi$ for the charged particles. The detector response was studied using a GEANT-based Monte Carlo (MC) simulation including radiation corrections~\cite{GEANTMC}.

The $\psi(2S)$ data are analyzed for the \textit{exclusive} decay, 
$\psi(2S) \to \pi^{+}\pi^{-} J/\psi$, $J/\psi \to \mu^{+}\mu^{-}$ and
for the \textit{inclusive} decay, $\psi(2S)\to K_S+X$,  
$K_S\to \pi^{+}\pi^{-}$. We select events with
 well-measured tracks by requiring that they be fully contained in the 
barrel region of the detector, $|\cos\theta(polar)|$ $<$ 0.8, and have transverse
 momenta $>120$~MeV.
 For the pions from $K_S$ decay, we make the additional requirement that 
they originate from a  common vertex displaced from the interaction point 
by more than 10 mm. We require a $K_S$ flight distance significance of more 
than three standard deviations.
 We accept $K_S$ candidates with mass in the range $497.7\pm12.0$ 
MeV.  
We identify muons from $J/\psi$ decays as having 
momenta more than 1 GeV, and $E_{CC}/p<0.25$ for at least one muon candidate,
and  $E_{CC}/p<$0.5 for the other muon, where $E_{CC}$ is the
 energy deposited in electromagnetic calorimeter associated with the track of
momenta $p$.

  We require that there should be only two identified 
pions and two identified muons with opposite charges in the event.
The momenta of $\mu^{+}\mu^{-}$ pairs is kinematically fitted to the 
KEDR $J/\psi$ mass, $M(J/\psi)_{\mathrm{KEDR}}=3096.917$~MeV,  
and only events with $\chi^2<20$ are
accepted. We also require that there should not be any isolated shower 
with energy more than 50 MeV in the event.

The $\psi(3770)$ data are analyzed for the decays 
$\psi(3770)\to D^0\bar{D^0}$, $D^0/\bar{D^0} \to K3\pi$.
We select $D^0$ candidates using the standard CLEO D-tagging criteria, 
which impose a very loose requirement on the beam energy constrained 
$D^0$ mass, as described in Ref.~\cite{dtag}.  
We again select well-measured tracks as described above, and in addition require that they have energy loss, $\mathrm{dE/dx}$, in the drift chamber consistent with the pion or kaon hypothesis within three standard deviations.

There are three distinct steps involved in our analysis:

\begin{enumerate}
\item Determination of the improved energy calibration of the CLEO--c detector for charged particles by using the \textit{exclusive} decay, $\psi(2S)\to J/\psi\pi^+\pi^-$, and the precision masses of $\psi(2S)$ and $J/\psi$.
\item Precision measurement of the mass of $K_S$ in the \textit{inclusive} decay, $\psi(2S)\to K_S+X$, $K_S\to\pi^+\pi^-$ using the improved calibration.
\item Precision measurement of the mass of $D^0$ in the \textit{exclusive} decay, $D^0\to K3\pi$ by monitoring and correcting for small changes in calibration as revealed by $M(K_S)$ determined for individual subruns.
\end{enumerate}

The first step consists of determining the new calibration for the momenta 
of charged particles with the highest possible precision. The charged particle energy calibration generally used in the analyses of CLEO--c data is based on tuning of the 
nominal magnetic  field of the CLEO III detector done in 2003. By requiring that in the decays 
$\psi(2S)\to \mu^{+}\mu^{-}$, and $J/\psi \to \mu^{+}\mu^{-}$ 
 the reconstructed  $\psi(2S)$ and  $J/\psi$  masses be equal to their then known average PDG2002 values of $M(J/\psi)=3096.87\pm0.04$~MeV and 
$M(\psi(2S))=3685.96\pm0.09$~MeV, it was determined that the nominal B-field of the solenoid needed to be multiplied by a default correction factor $\mathrm{B_{COR}(default)}=0.9952$. 
With the improved values of $M(\psi(2S))$ and $M(J/\psi)$ now available, and with our required level of high precision, it is necessary to determine the new value 
of the B-field correction factor appropriate for our present measurements.

\begin{figure}[!t]
\includegraphics*[width=3.5in]{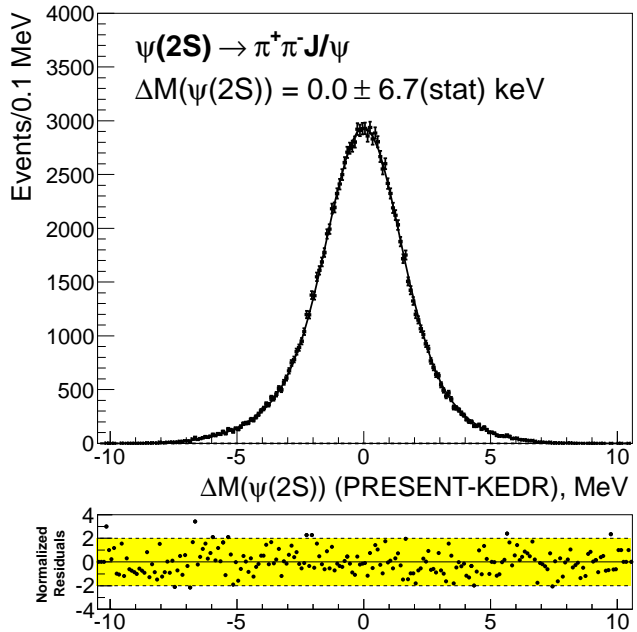}
\caption{  Results of the unbinned maximum likelihood fit to the 
 distribution $\Delta M(\psi(2S))\equiv M(\psi(2S))_{\mathrm{PRESENT}}-M(\psi(2S))_{\mathrm{KEDR}}$
for the exclusive decays $\psi(2S) \to \pi^{+}\pi^{-} J/\psi$, 
$J/\psi \to \mu^{+}\mu^{-}$ using the corrected magnetic field.}
\end{figure}

The KEDR determined precision values of the masses of the $M(J/\psi)$ and $M(\psi(2S))$ provide us the opportunity to determine 
precision calibration for charged pion momenta in the decays  $\psi(2S) \to \pi^{+}\pi^{-} J/\psi$. 
The pions in this decay have momenta up to $\sim$400 MeV, and the calibration 
obtained for them can be reliably used in the study of 
$D^0\to K3\pi$ decays which contain charged pions and kaons in a 
similar range of momenta. 
By analyzing our data for the exclusive reaction,  
$\psi(2S) \to \pi^{+}\pi^{-} J/\psi$, with $M(J/\psi)$ fixed at the precision value, 
 $M(J/\psi)_{\mathrm{KEDR}}$=3096.917 MeV we determine the new value of 
the solenoid B-field
correction factor,  $\mathrm{B_{COR}(new)}$, which corrects the pion momenta such 
the mass of $\psi(2S)$ we measure, $M(\psi(2S))_{\mathrm{PRESENT}}$ equals the precision value  
$M(\psi(2S))_{\mathrm{KEDR}}=3686.114$ MeV.  
It is found that the CLEO--c default value
$\mathrm{B_{COR}(default)}$ has to be increased by 0.0289\%, or 
2.89$\times$10$^{-4}$, so that $\mathrm{B_{COR}(new)}$=0.995488. The $\psi(2S)$
mass spectrum obtained with this corrected B-field is shown in Fig. 1 in
 terms of  $\Delta M(\psi(2S))\equiv M(\psi(2S))_{\mathrm{PRESENT}}-M(\psi(2S))_{\mathrm{KEDR}}$.
The unbinned spectrum is fitted with a constant linear background 
($\sim$2 counts/0.1 MeV bin) and a peak which is the sum of a simple Gaussian
function (54\%), and a bifurcated Gaussian function (46\%)  with the same mean.
 The fit leads to $N(\psi(2S))$=125,299$\pm$354 events, FWHM=4.4 MeV, 
$\Delta M(\psi(2S))$=0.0$\pm$6.7 keV(stat), and $\chi^2/d.o.f$=0.85.
 The normalized residuals for the
fit defined as  $[N(observed)-N(fit)]/\sqrt{N(fit)}$, are also shown. 
All subsequent spectra in this paper are fitted in 
the same manner.

The second step of analysis consists of a precision determination of $M(K_S)$ , the mass of the $K_S$ 
meson which we use to monitor the stability of the magnetic field for the different
$\psi(3770) \to D \bar{D}$ data subruns. We analyze the 
 \textit{inclusive} reaction  
$\psi(2S)\to K_S+X$ to determine $M(K_S)$. 
We use the precision calibration of the B-field as determined in the first step
for this purpose. The pions in $\psi(2S) \to \pi^{+}\pi^{-} J/\psi$ 
calibration have momenta up to 400 MeV. 
For determining $M(K_S)$ we only use $K_S$ with momenta $p(K_S)<$ 400 MeV for which 95\% 
of pions from $K_S\to \pi^{+}\pi^{-}$ decay have momenta $<$360 MeV. 
Fig. 2 shows that for $K_S$ of momenta $<$400 MeV, the $\pi^{+}$ and $\pi^{-}$ from $\psi(2S)$ decay
and from $K_S\to \pi^{+}\pi^{-}$ decay have nearly identical pion 
momentum distributions and angular distributions of the pions with respect 
to beam. 

\begin{figure}[!t]
\includegraphics*[width=3.in]{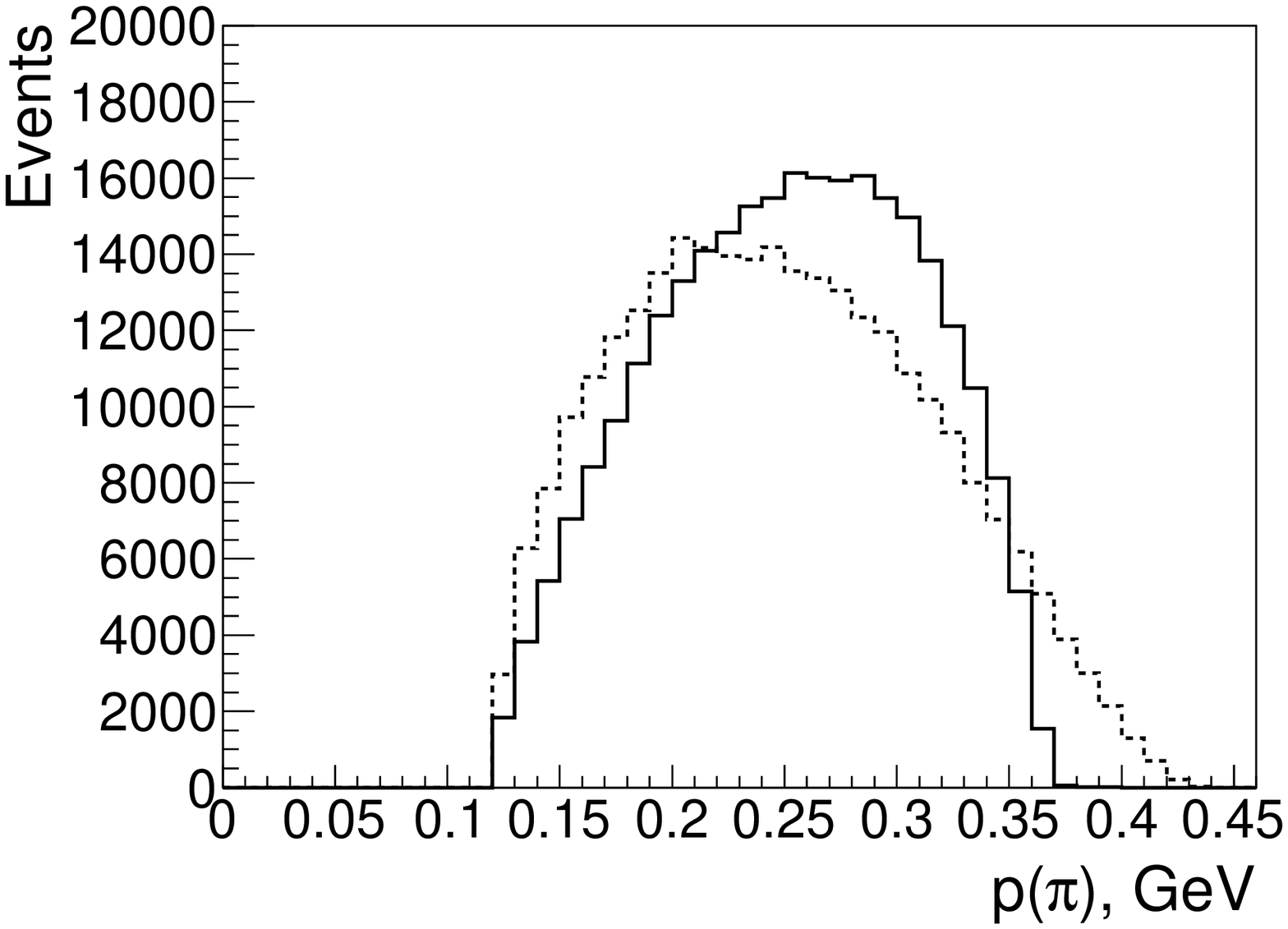}
\includegraphics*[width=3.in]{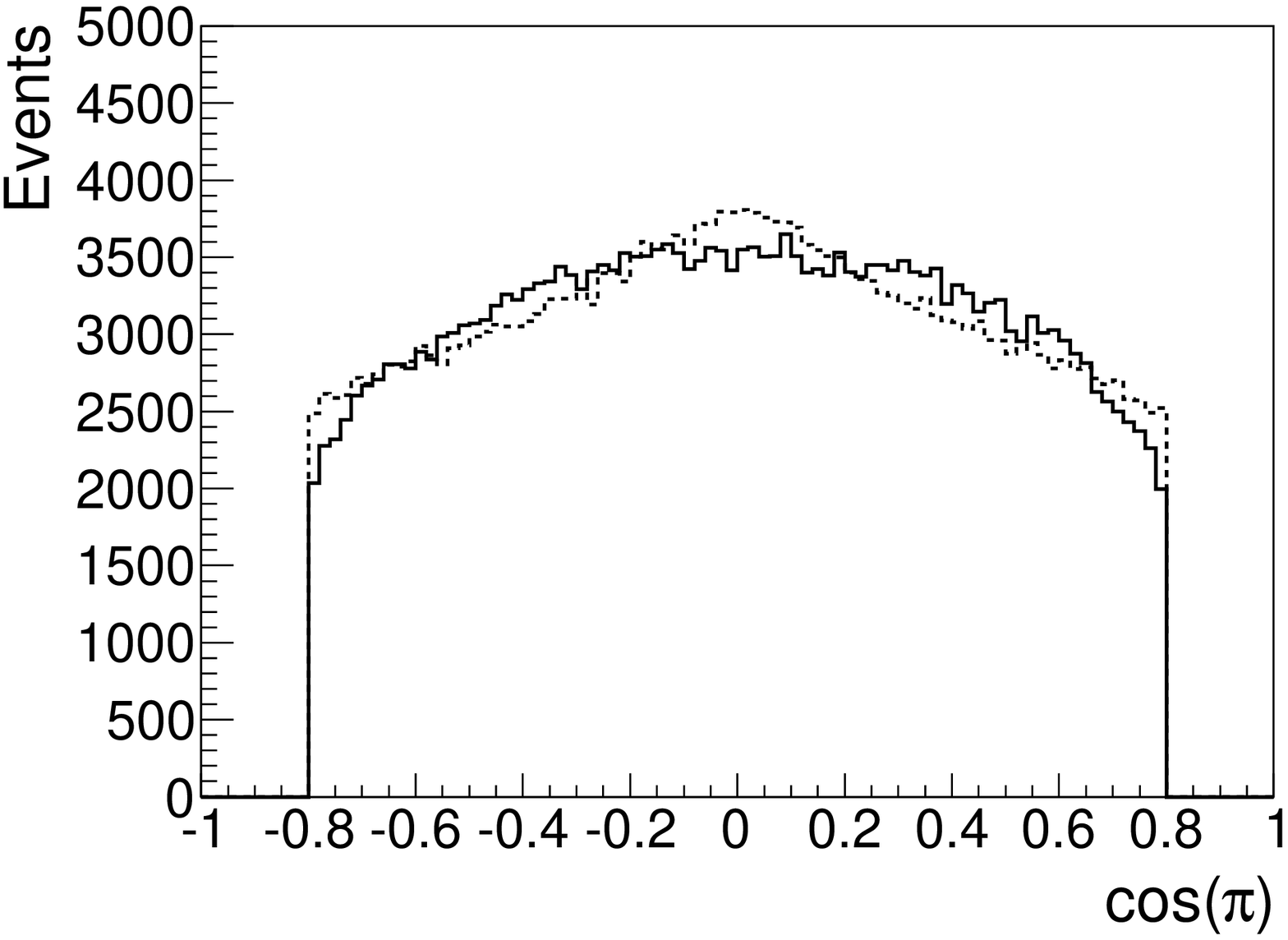}
\caption{Comparison of momentum distributions (upper) and angular
distributions (lower) for pions in $\psi(2S) \to \pi^{+}\pi^{-} J/\psi$ 
decays (full histogram) and for pions in  $K_S\to \pi^{+}\pi^{-}$ decays (dashed histogram).}
\end{figure}

Fig. 3 shows the $M(\pi^{+}\pi^{-})$ distribution corresponding to 
$\mathrm{B_{COR}(new)}$=0.995488 for events from
the decay $K_S\to \pi^{+}\pi^{-}$. 
The distribution is fitted as described before, 
with the fraction of the simple Gaussian and bifurcated 
Gaussian being 52\% and 48\%, respectively. It leads to 
$N(K_S)=261,394\pm$752, FWHM=4.1 MeV, $\chi^2/d.o.f.$=1.2, and
\begin{equation} 
M(K_S)_{\mathrm{PRESENT}}=497.607\pm0.007\mathrm{(stat)~MeV}.
\end{equation}

\begin{figure}[!t]
\includegraphics*[width=3.5in]{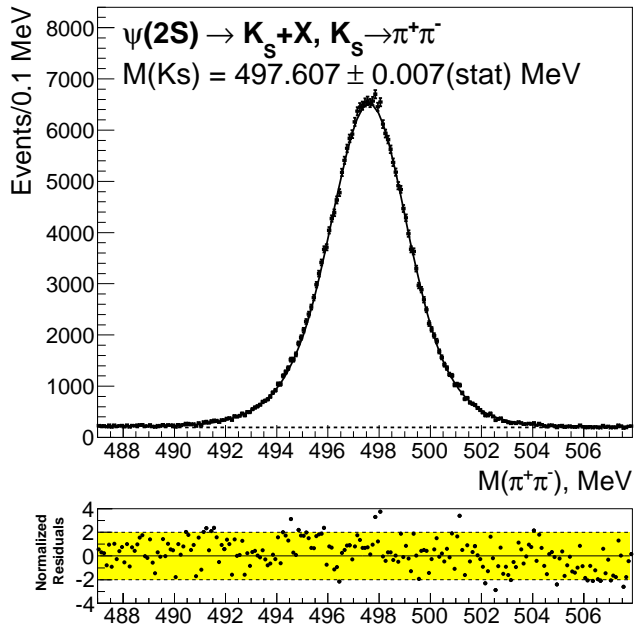}
\caption{ Results of the unbinned maximum likelihood fit to the 
invariant mass distribution $M(\pi^+\pi^-)$ for the inclusive reaction
$\psi(2S)\to K_S+X$,  $K_S\to \pi^{+}\pi^{-}$, using the corrected magnetic 
field.} 
\end{figure}

\begin{table}[htb]
\caption{Illustrating stability of $D^0$ mass for different ranges of kaon and pion momenta.}
\begin{ruledtabular}
\begin{tabular}{ccc}
$p(K,\pi$'s), MeV      &  $N(D^0)$   & $M(D^0)$, MeV   \\
\hline
$<$600             &50,964$\pm$316        &$1864.849\pm0.027$ \\
$<$650             &62,557$\pm$361        &$1864.845\pm0.025$ \\
$<$700             &69,461$\pm$383        &$1864.849\pm0.024$ \\
$<$750             &73,046$\pm$404        &$1864.847\pm0.023$ \\
$<$800             &74,728$\pm$412        &$1864.846\pm0.022$ \\
\end{tabular}
\end{ruledtabular}
\end{table}

Although we have used $M(\psi(2S))$ based energy calibration obtained for 
pions with $p(\pi)<400$~MeV to determine $M(K_S)$ for $K_S$ decays 
with $p(K_S)<400$~MeV, we find that the calibration is good for 
higher momenta. For example, we find that if decays with $p(K_S)$ up 
to 650~MeV are included, $M(K_S)$ varies by less than 1$\sigma$, or $<10$~keV.

The third step of analysis consists of the determination of the mass of the $D^0$ meson using the $\psi(3770)\to D^0\bar{D^0}$ data taken at $\psi(3770)$, $\sqrt{s}=3770$ MeV, and reconstructing $D^0$ in the decay $D^0\to K3\pi$. 
The data which we analyze were taken in four subruns totaling 
580 pb$^{-1}$. These data were taken after a three months shut--down after 
$\psi(2S)$ running of CLEO/CESR. 
Before analyzing the $D^0\to K3\pi$ decays, it is necessary to determine the
appropriate  $\mathrm{B_{COR}}$ values for the $D^0$ subruns.
We do so by analyzing each individual subrun for the \textit{inclusive} decay, $D\to K_S+X,~K_S\to\pi^+\pi^-$, with $p(K_S)<650$~MeV and determining individual values of $\mathrm{B_{COR}}$ required to make $M(K_S)$ equal to $M(K_S)_{\mathrm{PRESENT}}$, as determined in the second step. 
More than 99\% of the pions in the inclusive decay $D\to K_S+X,~K_S\to\pi^+\pi^-$ 
have momenta $<$650 MeV for which our calibration of pion momenta 
is appropriate. 
The correction factors for  $\mathrm{B_{COR}(default)}$ for individual 
subruns so determined are found 
to be  $(0.79,0.49,0.68,0.26)\times10^{-4}$. These are smaller than  
$2.89\times10^{-4}$
 determined in the first step by fitting $M(\psi(2S))$ data taken before the 
three months shut--down. Using the above
 individual $\mathrm{B_{COR}}$ values the invariant mass of $D^0$ was reconstructed for the decay $D^0\to K3\pi$ for each of the four subruns. Their weighted 
average is 
\begin{equation}
<M(D^0)>=1864.833\pm0.024\mathrm{(stat)~MeV}.
\end{equation}

\begin{figure}[!t]
\begin{center}
\includegraphics[width=3.5in]{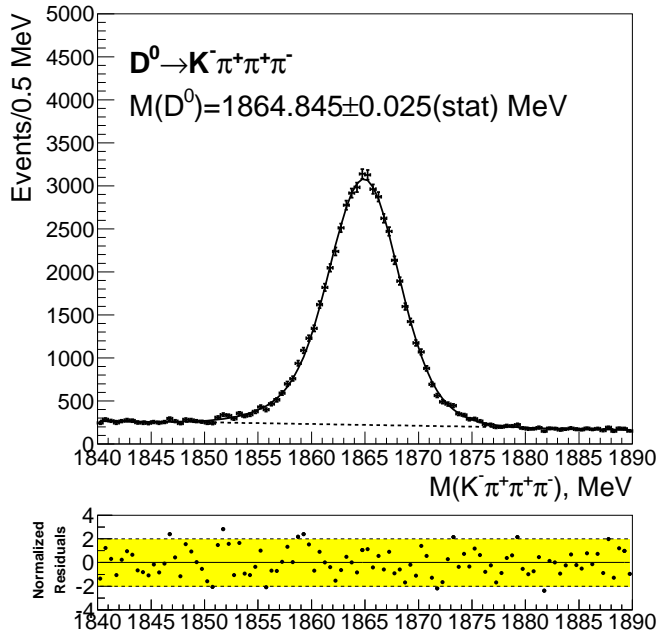}
\end{center}
\caption{Invariant mass spectra for the decays
$D^0\to K^{-}\pi^{+}\pi^{+}\pi^{-}$ (plus  charge conjugation decays).}
\end{figure}

\begin{figure}[!t]
\includegraphics*[width=3.in]{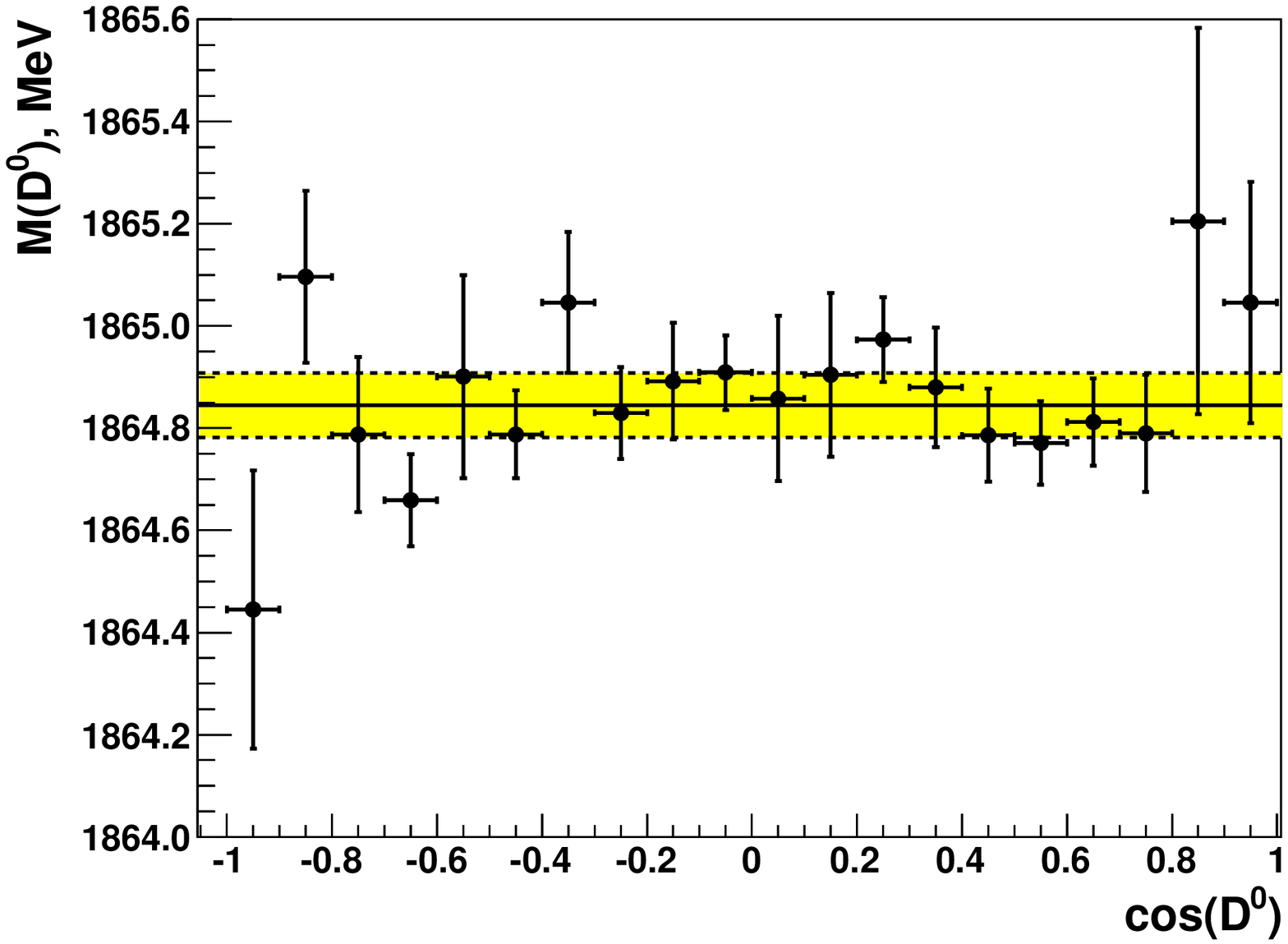}
\includegraphics*[width=3.in]{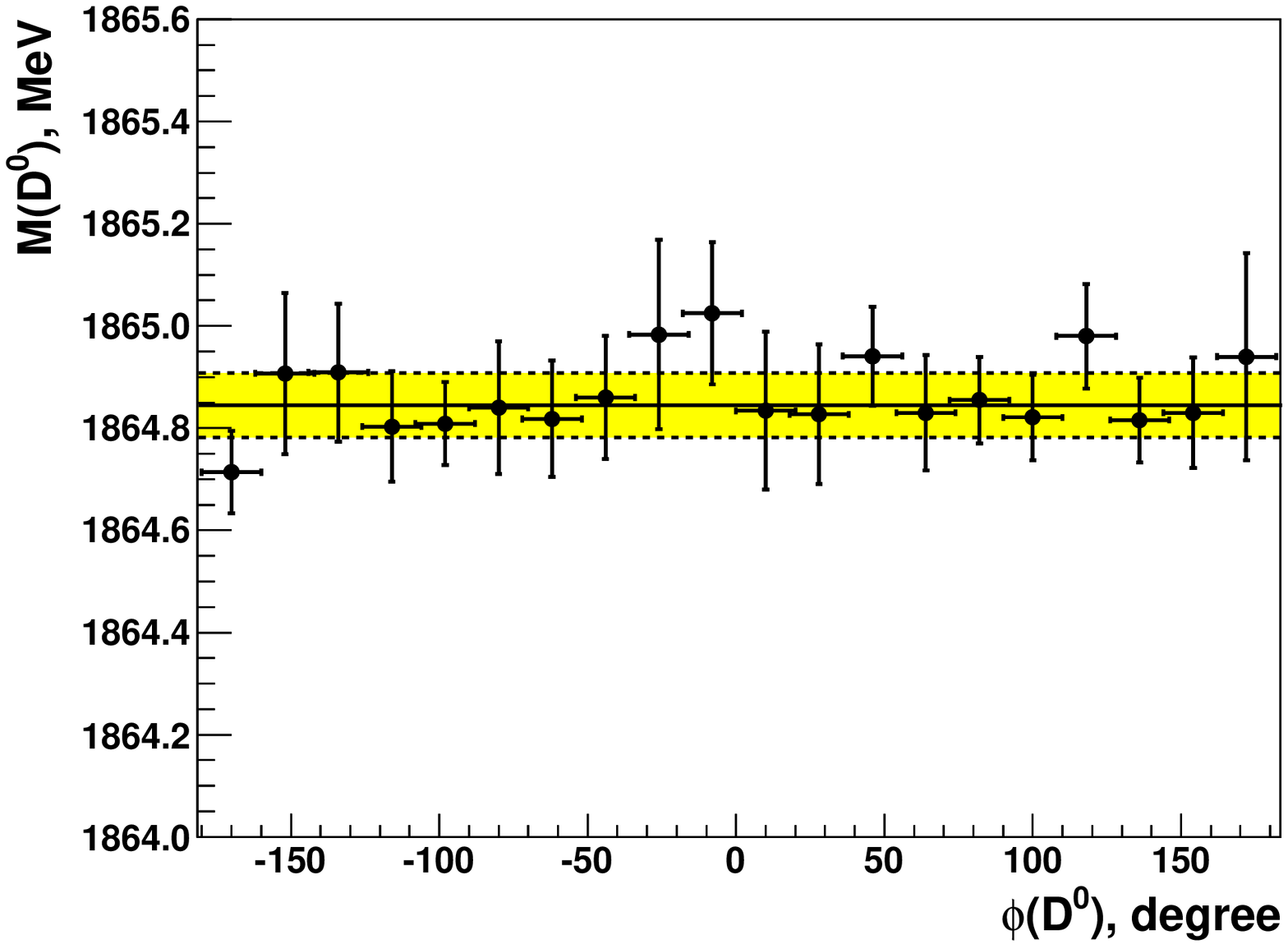}
\caption{Values of $M(D^0)$ obtained from fits to data divided into subsets 
in $\cos \theta$ and $\phi$
of $D^0$ mesons. Solid and dashed lines correspond to the central value and total error band corresponding to our present measurement 
$M(D^0)=1864.845\pm0.063~\mathrm{MeV}.$} 
\end{figure}

\begin{figure*}[!tb]
\begin{center}
\includegraphics[width=3.3in]{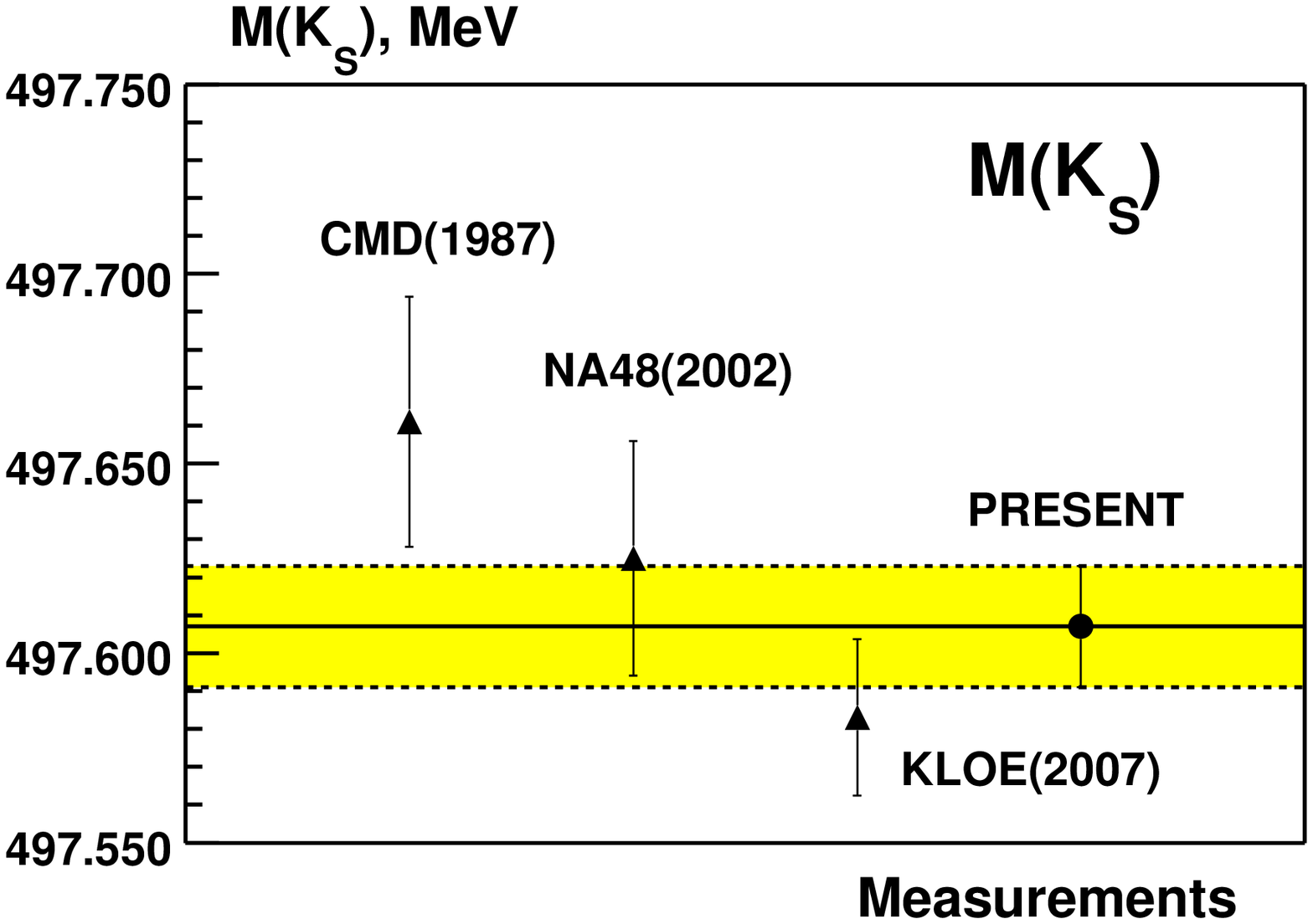}
\includegraphics[width=3.3in]{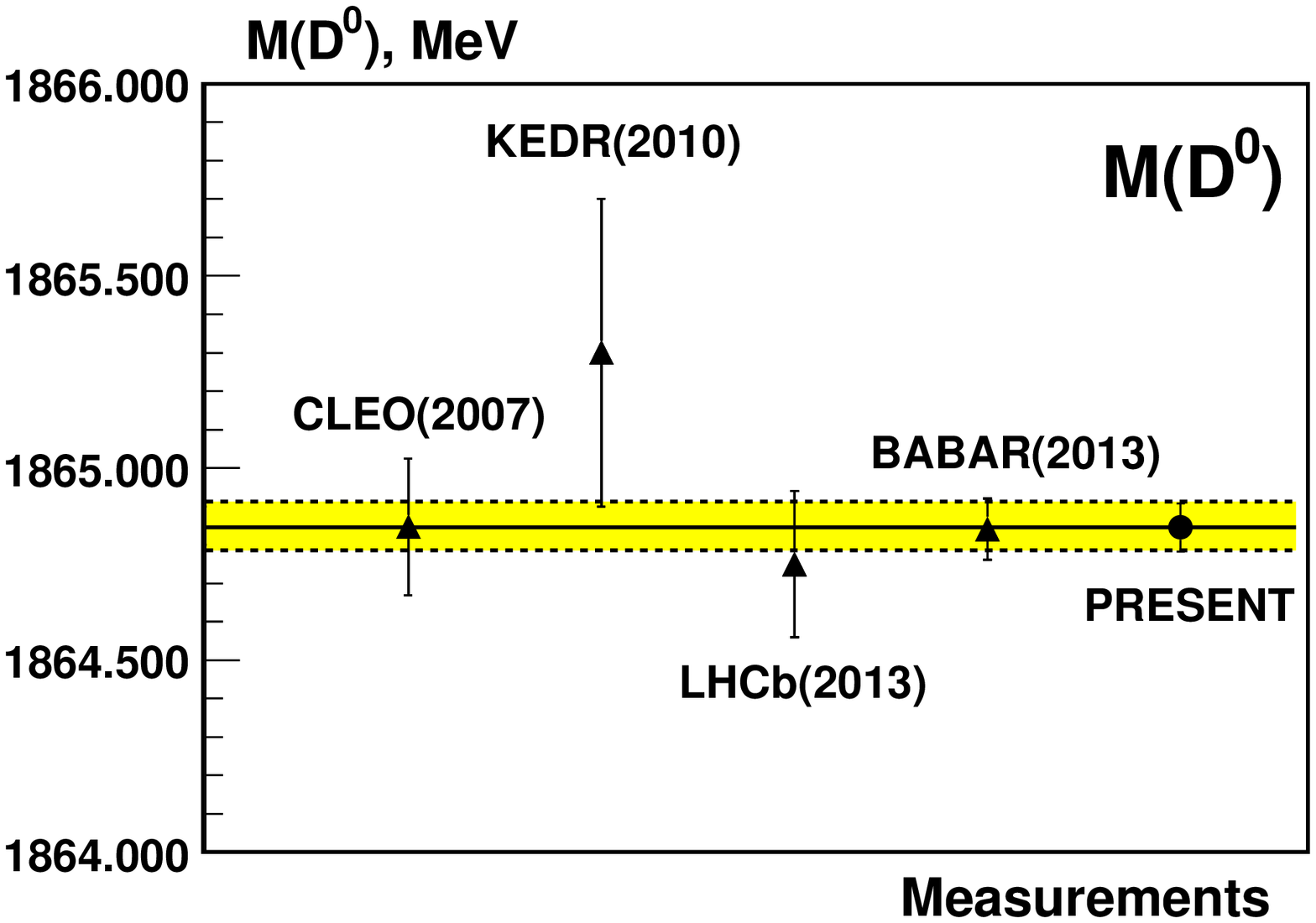}
\end{center}
\caption{Summary of $M(K_S)$ and $M(D^0)$ from individual measurements from different experiments. The statistical and systematic errors of the measurements were added in quadrature. Solid and dashed lines correspond to the central value and total error band corresponding to our present measurement $M(K_S)=497.607\pm0.016~\mathrm{MeV},~M(D^0)=1864.845\pm0.063~\mathrm{MeV}$. }
\end{figure*}

For our final result we sum the corrected spectra for the four subruns and
 fit the summed spectrum as described before.  The  fractions of the simple 
Gaussian function and the bifurcated Gaussian function are 67\% and 33\%, 
respectively,  The results of the fit shown
in Fig.~4 are $N$=62,557$\pm$361 events, FWHM=8.9 MeV,  
$\chi^2/d.o.f.$=0.91, and

\begin{equation}
M(D^0)=1864.845\pm0.025\mathrm{(stat)}~\mathrm{MeV}.
\end{equation}

Table I illustrates that $M(D^0)$ is stable to within $\pm5$~keV even for $\pi$ and $K$ momenta up to 800~MeV.

\begin{table}[htb]
\caption{Systematic uncertainties in $M(K_S)$.}
\begin{ruledtabular}
\begin{tabular}{lc}
Source: variation & Uncertainty in $M(K_S)$, keV   \\
\hline
$\psi(2S)$ mass: -18+13 keV   & -12.3+8.9    \\
$J/\psi$ mass: $\pm$12 keV &  8.2    \\
Fit Range width, $\pm2$~MeV    & 4    \\
Background polynomial, 1,2 order    & 1  \\
$\psi(2S)$ formation energy     & 5  \\
Total   & 15  \\
\end{tabular}
\end{ruledtabular}
\end{table}

The systematic uncertainties in $M(K_S)$ and $M(D^0)$ were obtained as follows.

For $M(K_S)$ measurement, we have corrected the magnetic field using KEDR measured $M(\psi(2S))$ and $M(J/\psi)$, which have the total errors 
of $-18+13$~keV and $\pm12$~keV, respectively~\cite{kedr,kedr1}. The 
change in $M(K_S)$ due to the change in the magnetic field is 
found to vary linearly with the change in $M(\psi(2S))$ and $M(J/\psi)$, and 
is a factor 1.46  smaller.
We therefore assign $\sim$$_{-12.3}^{+8.9}$~keV, and $\sim\pm8.2$~keV, as the uncertainties in $M(K_S)$ due to the uncertainties in 
 $M(\psi(2S))$ and $M(J/\psi)$, respectively.
The variation of the fit range by $\pm$2 MeV yields a change of 
$\pm$4 keV in $M(K_S)$.
Changing the fits to the background from polynomials of order one to polynomials of order two changes $M(K_S)$ by $<1$~keV. 
The effect of the possible formation of $\psi(2S)$ at an energy different
from $M(\psi(2S))_{\mathrm{KEDR}}$ was investigated in detail.
The uncertainty in the formation energy  was estimated by fitting  the 
$\psi(2S)$ mass distribution with MC shape using different beam energies, and was found to be $\pm7$~keV.
It contributes $\pm5$~keV to the
systematic uncertainty in $M(K_S)$. 
The systematic uncertainties in $M(K_S)$ are listed in Table II.
The sum in quadrature of all the above $\mathrm{B_{COR}}$ 
contributions is a total systematic uncertainty of $\pm$15~keV.

We have studied $K_S$ mass dependence on momenta, polar angle  $\theta$ 
 and azimuthal angle $\phi$ of $K_S$ with respect to the positron beam. 
The $K_S$ mass values in all cases are seen to be statistically in agreement 
with the average value, with 
$\chi^2/d.o.f.$ equal to 0.76, 0.79, 0.96 for momenta,  $\cos\theta$, and
$\phi$.

Our final result for $M(K_S)$ is thus
\begin{equation}
M(K_S)_{\mathrm{PRESENT}}=497.607\pm0.007(\mathrm{stat})\pm0.015(\mathrm{syst})\;\mathrm{MeV}.
\end{equation} 

\begin{table}[htb]
\caption{Systematic errors in $M(D^0)$ for the range of variation of different parameters.}
\begin{ruledtabular}
\begin{tabular}{lc}
Source: variation      &  Uncertainty in $M(D^0)$, keV   \\
\hline
$|\cos\theta(polar)|_{\mathrm{max}}$: 0.8, 0.75    & 6    \\
$p_{\mathrm{min}}$(trans): 120, 135~MeV   & 6    \\
$p_{\mathrm{max}}$(total): 650, 550~MeV   & 15    \\
Fit Range width, $\pm5$~MeV    & 12    \\ 
Background polynomial 1,2 order    & 4  \\
MC Input/Output of $M(D^0)$    & 7  \\
Total: event selection and fit    & 22 \\
\hline\hline
Error in $K_S$ mass: $\pm$16 keV  & 52 \\
Error in $K^{\pm}$ mass: $\pm$16 keV      & 12  \\
Total: kaon masses   & 53  \\
\end{tabular}
\end{ruledtabular}
\end{table}

The systematic errors in $M(D^0)$ are listed in Table III. They are 
dominated by uncertainties in the masses of the kaons. 
The $\pm$16$~\mathrm{keV}$ uncertainty in the mass of $K_S$ leads to 
the largest uncertainty, $\pm$52$~\mathrm{keV}$ in $M(D^0)$. 

The PDG(2012) mass of $K^\pm$ has an error of 
$\pm16$~keV~\cite{pdg}. It leads to $\pm$12~keV uncertainty in $M(D^0)$, 
 which is calculated by changing of $M(K^\pm)$ by  $\pm$16 keV.
Added in quadrature, the total systematic uncertainty due to uncertainties 
in kaon masses is $\pm$53$~\mathrm{keV}$.

Other contributions to systematic error in $M(D^0)$ due to event selection 
and peak fitting procedure are all smaller,  
as shown in Table III. They include variation of maximum value of 
$|\cos\theta|$ for decay particles, variation of minimum value of 
transverse momenta and 
maximum value of total momenta of all particles, and variation of the fit 
range and background shape. We estimate the uncertainty in our analysis
procedure as the difference between MC input and output values of $M(D^0)$.
The difference is found to be $\Delta M(D^0)$(output--input)=7$\pm$1 keV and
we assign a systematic uncertainty of $\pm$7 keV. Added in quadrature, the total 
systematic uncertainty due to event selections and fit procedure is 
$\pm$22 keV.

In Fig. 5 we show the $D^0$ mass difference dependence  
on $\cos\theta$ and azimuthal angle $\phi$. 
All $M(D^0)$ values are found to be statistically in agreement with the 
 average value, with $\chi^2/d.o.f.$ of 0.96 and 0.47 for  
$\cos\theta$ and $\phi$, respectively.

Thus our final result for $M(D^0)$ is
\begin{eqnarray}
\nonumber M(D^0)_\text{PRESENT}= & 1864.845\pm0.025\mathrm{(stat)}\pm0.022\mathrm{(syst)} \\
   & \pm0.053(\mathrm{kaon~masses})~\mathrm{MeV}.
\end{eqnarray}
With all uncertainties added in quadrature, our present results are
\begin{eqnarray}
M(K_S)_{\mathrm{PRESENT}}=497.607\pm0.016~\mathrm{MeV},\\
M(D^0)_{\mathrm{PRESENT}}=1864.845\pm0.063~\mathrm{MeV}.
\end{eqnarray}
Both $M(K_S)$ and $M(D^0)$ are presently the world's most precise single measurements of these masses.
Our $M(D^0)$ agrees with our previous measurement~\cite{d0pub}, and has a factor three smaller uncertainty. It also agrees with the recent the BaBar 
result~\cite{babar}, and is based on fourteen times larger number of events, 
has factor two smaller statistical error, and $\sim20\%$ smaller overall error.
Fig.~6 shows these results together with results of previous mass 
measurements~\cite{cmd,na48,kloe,Trilling,Barlag,d0pub,kedr3,lhcb,babar}. 
The world average of all measurements, determined mainly by our results in 
Eq. 7, and
the BaBar results $M(D^0)$=1864.841$\pm$0.079 MeV~\cite{babar}, is
$M(D^0)$=1864.843$\pm$0.044 MeV. The 1992 CLEO measurement~\cite{cleodm}, 
adopted by PDG~\cite{pdg}, gives $M(D^{*0})-M(D^{0})$=142.12$\pm$0.07 MeV.
Thus, $M(D^{*0})$=2006.963$\pm$0.083 MeV, and  
$M(D^{0})+M(D^{*0})$=3871.806$\pm$0.112 MeV.
This leads to the binding energy of  X(3872), 
B.E.~X(3872)=(3871.806$\pm$0.112)--(3871.68$\pm$0.17)=0.126$\pm$0.204 MeV.

This investigation was done using CLEO data, and as members of 
the former CLEO Collaboration we thank it for this privilege.
This research was supported by the U.S. Department of Energy.

\end{document}